# MULTI-WAVELENGTH MONITORING OF THE CHANGING-LOOK AGN NGC 2617 DURING STATE CHANGES

V.L.Oknyansky [1], C.M.Gaskell [2], N.A.Huseynov [3], Kh.M.Mikailov [3], V.M.Lipunov [1,4], N.I.Shatsky [1], S.S.Tsygankov [5], E.S.Gorbovskoy [1], A.M.Tatarnikov [1], V.G.Metlov [1], K.L.Malanchev [1], M.B.Brotherton [6], D.Kasper [6], P.Du [7], X.Chen [8], M.A.Burlak [1], D.A.H.Buckley [9], R.Rebolo [10], M.Serra-Ricart [10], R.Podesta [11], H.Levato [12]

[1] M.V.Lomonosov Moscow State University, Sternberg Astronomical Institute, Moscow, Russian Federation, *oknyan@mail.ru*

[2] Department of Astronomy and Astrophysics, University of California, Santa Cruz, USA

[3] Shamakhy Astrophysical Observatory, National Academy of Sciences, Pirkuli, Azerbaijan

[4] M.V. Lomonosov Moscow State University, Physics Department, Moscow, Russian Federation

[5] Tuorla Observatory, Department of Physics and Astronomy, University of Turku, Piikkiö, Finland

[6] University of Wyoming, Laramie, USA

[7] Institute of High Energy Physics, Chinese Academy of Sciences, Beijing, China

[8] School of Space Science and Physics, Shandong University , Shandong, China

[9] The South African Astronomical Observatory, Observatory, South Africa

[10] The Instituto de Astrofisica of Canarias, Tenerife, Spain

[11] OAFA, National University of San Juan, San Juan, Argentina

[12] ICATE, National University of San Juan, San Juan, Argentina

ABSTRACT. Optical and near-infrared photometry, optical spectroscopy, and soft X-ray and UV monitoring of the changing-look active galactic nucleus NGC 2617 show that it continues to have the appearance of a type-1 Seyfert galaxy. An optical light curve for 2010–2017 indicates that the change of type probably occurred between 2010 October and 2012 February and was not related to the brightening in 2013. In 2016 and 2017 NGC 2617 brightened again to a level of activity close to that in 2013 April. However, in 2017 from the end of the March to end of July 2017 it was in very low level and starting to change back to a Seyfert 1.8. We find variations in all passbands and in both the intensities and profiles of the broad Balmer lines. A new displaced emission peak has appeared in Hβ. X-ray variations are well correlated with UV–optical variability and possibly lead by ~2–3 d. The *K* band lags the *J* band by about 21.5 ± 2.5 d and lags the combined *B* + *J* bands by ~25 d. *J* lags *B* by about 3 d. This could be because J-band variability arises predominantly from the outer part of the accretion disc, while *K*-band variability is dominated by thermal re-emission by dust. We propose that spectral-type changes are a result of increasing central luminosity causing sublimation of the innermost dust in the hollow bi-conical outflow. We briefly discuss various other possible reasons that might explain the dramatic changes in NGC 2617.

## 1. Introduction

NGC 2617 is a typical representative of the objects called “Changing Look” AGNs (CL AGNs), that is, active galactic nuclei changing their spectral type (their Seyfert type). Before 2013, NGC 2617 received relatively little attention, since significant variability was not detected, and optical spectra were obtained only twice: in 1994 (Moran, Halpern & Helfand, 1996) and in 2003 (the 6dFGS spectrum). According to these spectra NGC 2617 could be classified as Sy 1.8. In 2013 spectra showed a dramatic change in the profiles of the Balmer lines compared with the 2003 spectra and NGC 2617 showed a type 1 Seyfert spectrum (Shappee et al., 2013). Since then NGC 2617 has been intensively studied (Shappee et al., 2014; Fausnaugh et al., 2017) by photometric and spectral monitoring during 2013-2014. In 2016 we started our monitoring as a continuation of the monitoring of Shappee et. al., but we carried out not only new photometric observations at X-ray, UV, optical, IR, but also processed MASTER archival observations from 2010 and archival *Swift* observations from 2013. The results of our multi-wavelength monitoring up to June 2016 are given in Oknyansky et al. (2017). We have continued our multi-wavelength monitoring of NGC 2617 from September 2016 to July 2017. We present here preliminary new results from this.

## 2. Observations

As in Oknyansky et al (2017; see details there), our new observational data includes IR *JHK* photometry (IR-camera Astronircam (Nadjip et al., 2017) of 2.5-m telescope of the SAI Caucasus Mountain Observatory), optical photometry (MASTER Global Robotic Net (Lipunov, 2010; Kornilov, 2012), AZT-5, Zeiss-600 ShAO, *Swift/UVOT*), UV photometry (*Swift/UVOT*), X-ray observations in the 0.3–10 keV band (S*wift/XRT*), optical spectrophotometry with the 2-m Zeiss telescope of the ShAO. In the 2016-2017 season, additional CCD photometric observations were added from a Zeiss-600 telescope with the Apogee Aspen CG42 CCD camera system at the Crimean Station of the SAI MSU, the 1-m telescope at Weihai Observatory China (see details, for example, Guo et al., 2014), and also for optical spectroscopy from the 2.3-meter WIRO telescope (for details see: *physics.uwyo.edu/~WIRO/Longslit/long_slit.html*).

## 3. Multi-wavelength light curves

Fig. 1 shows multi-wavelength light curves. The *B*-band magnitudes were obtained with different telescopes (see figure caption) and have not yet been reduced to a common system to remove small systematic differences.

One can see that there are fast components of variability with several brightenings and a slow decline in brightness. From the end of the March 2017 until the end of the observing season at the end of the July NGC 2617 was in very low state with a small amplitude of UV and optical variability. The correlation of the variability in different wave bands can be clearly seen as well as some difference in UV/optical and X-ray relative amplitudes of the last brightening. We used the same method for determining time delays as in our previous papers (see details in Oknysnaky et al., 2017). Our cross-correlation analysis confirms the high degree of correlation between the UV and X-ray variability but with a small lag about 3 days.

The new IR *JHK* data are in agreement with our previously determined lags between the optical and IR (Oknyansky et al., 2017). Fig. 2 shows UVW1 (*Swift*) variability compared with the *K*-band variability shifted by 24 days and with the amplitude scaled to the UVW1 variability. More details on our cross-correlation analysis will be given in a future paper.

## 4. Spectroscopic observations

In the 2016-2017 observational season we procured spectra of NGC2617 in the Hα and Hβ regions with 2 instruments: the 2.3-m WIRO telescope (8 dates) and 2-m ShAO telescope (7 dates). The times of spectroscopic observations are shown by arrows at the top of Fig. 1.

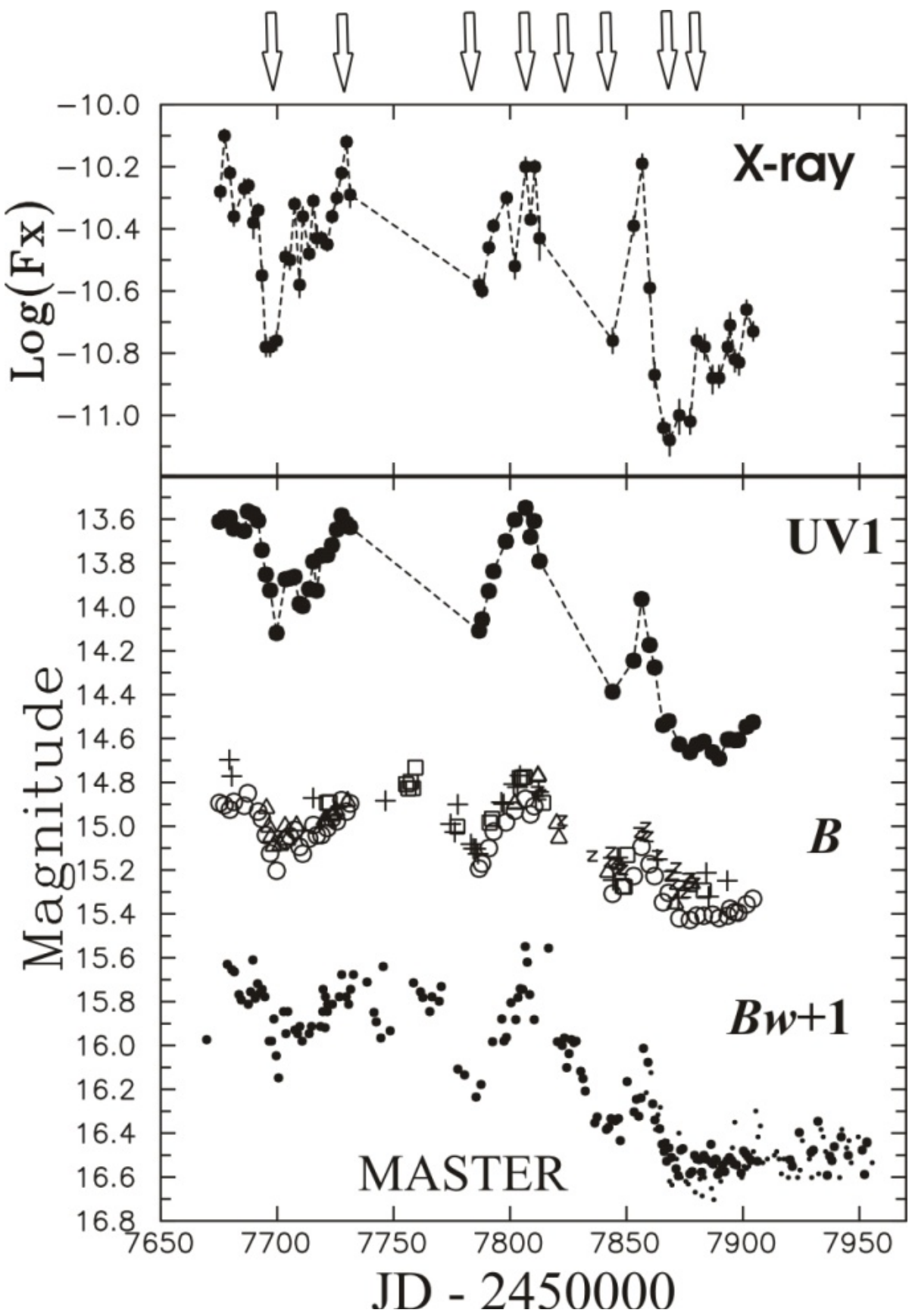

Figure 1: Multiwavelength light curves for NGC 2617. The top panel shows the log of the X-ray flux in $ergs/cm^2/s$ obtained with the *Swift*/XRT. The bottom panel shows (from top to bottom) UV1 = UVW1 magnitudes obtained with *Swift/UVOT*; ***B*** = *B* magnitudes obtained with the AZT-5 (+) and the Zeiss-600 (z) of the Crimean Station of the SAI, the Zeiss-600 of ShAO (squares), the 1-m telescope of Weihai Observatory (triangles), and from the *Swift* (open circles); $\boldsymbol{B_w}$ **+ 1** = unfitered ***W*** data data obtained with MASTER-SAAO and MASTER – IAC (filled circles), and MASTER-OAFA (points) reduced to the ***B*** band and shifted on $1^m$ for clarity. Arrows show times of spectroscopic observations

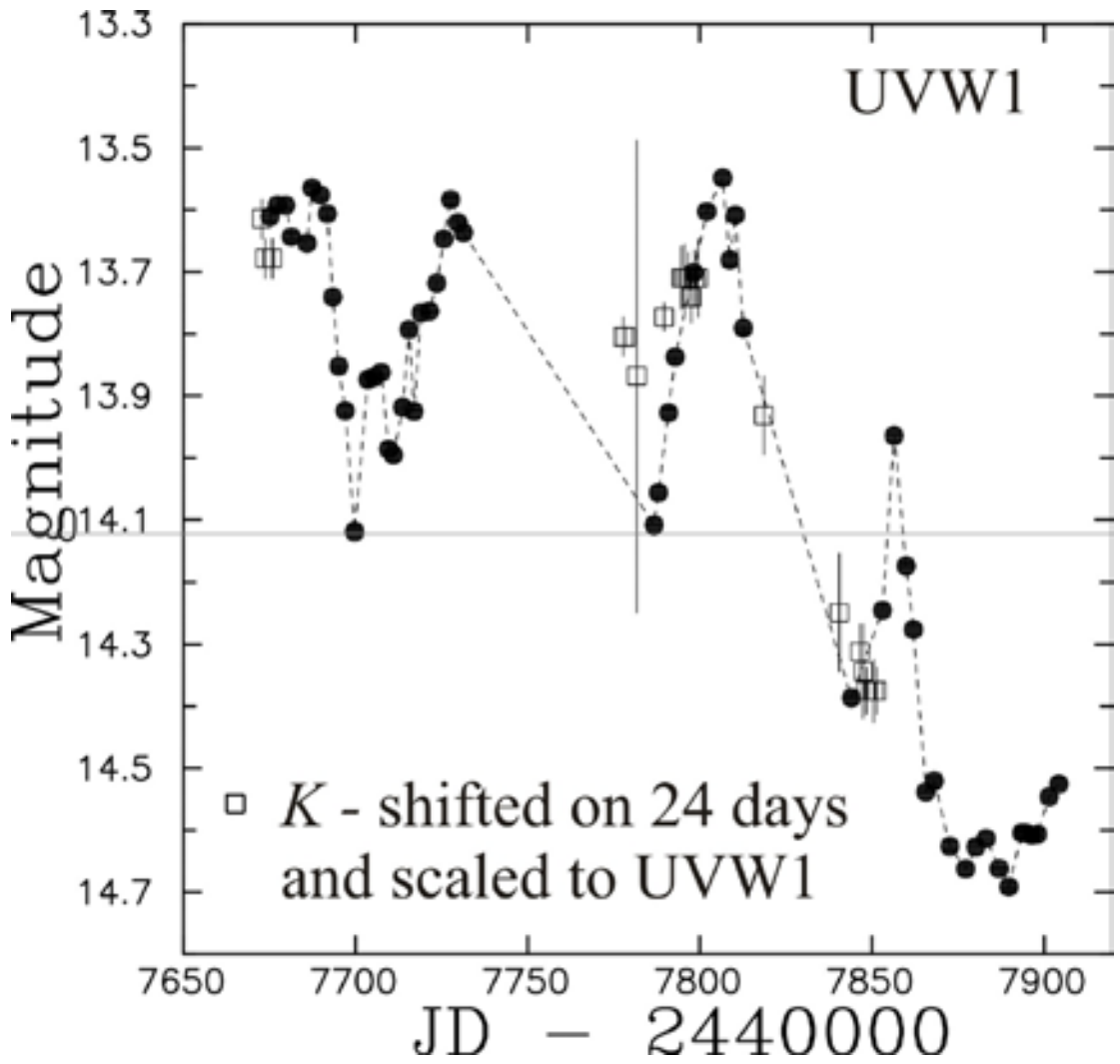

Figure 2: Light curves: points – **UVW1** (*Swift/UVOT*) and squares – *K* (2.5-m telescope of CMO SAI) shifted on -24 days and scaled to UVW1 following a linear regression.

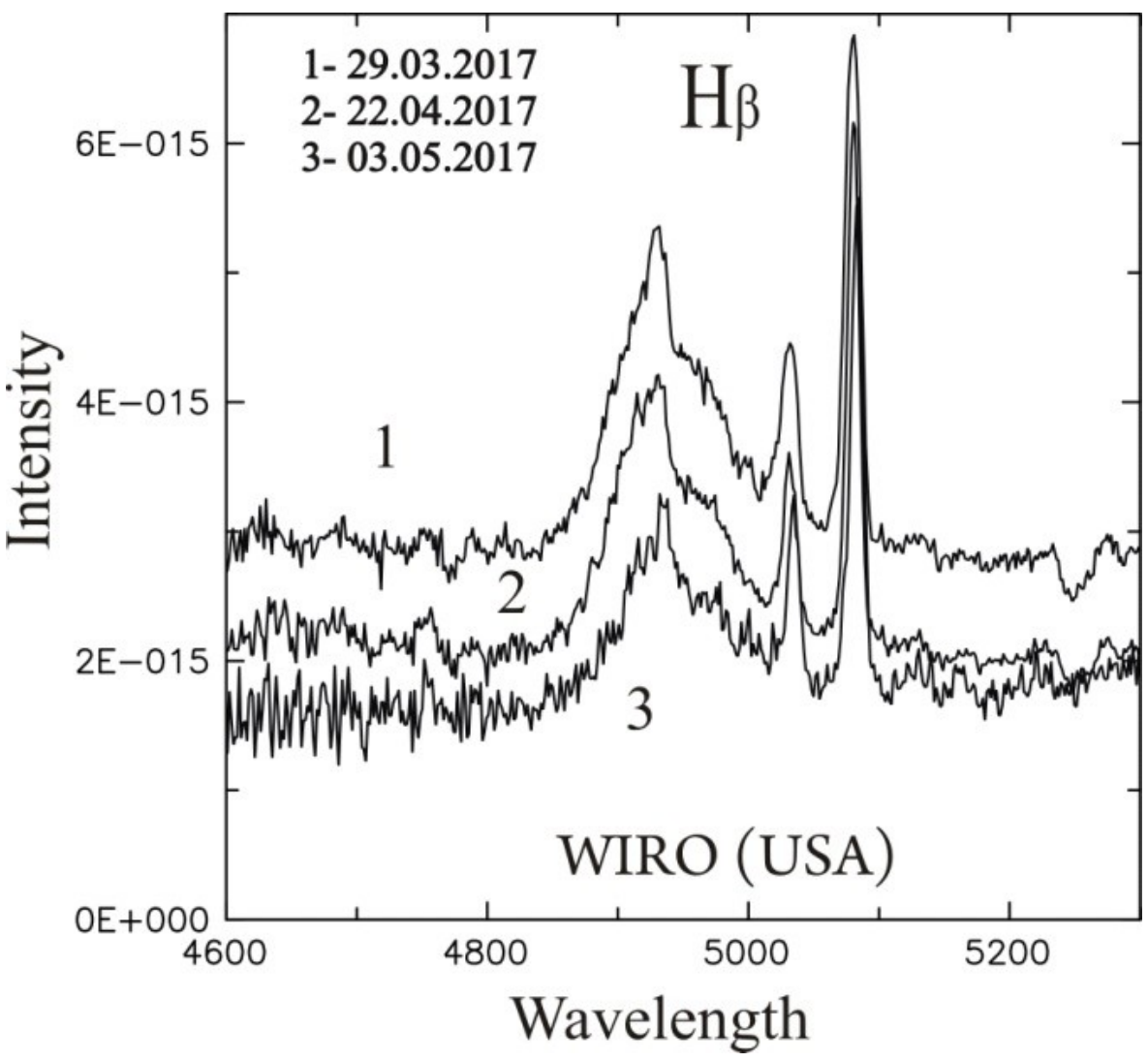

Figure 3: Calibrated spectra of NGC 2617 in Hβ region for 3 dates obtained with 2.3-m WIRO telescope. The intensity is in erg/s/sm$^2$/Å, the wavelength in Å.

For each date we usually obtained 3 spectra in Hα and Hβ regions and averaged the calibrated spectra for each date. Details will be described in a forthcoming paper. In Fig. 3 we show as an example just 3 of our spectra in the Hβ region obtained in WIRO. These spectra show how the intensity of Hβ was decreased from 2017 March 29 to 2017 May 3. From our photometry the deepest minimum at all wavelengths was at April 27-28. As was shown by Fausnaugh et al. (2017) the delay between the optical continuum and Hβ variability is about 6 days. This is in a good agreement with our results. In our last spectra Hβ still has a very low broad component but we suspect that NGC 2617 is going to change its spectral type back to a Seyfert 1.8 very soon.

## 5. Discussion

The most important question about CL AGNs is why they show the changes in luminosity and Seyfert type. We believe that the reason for the significant change in the luminosity of NGC 2617 is probably not a change in absorption alone. We propose that a change in the energy-generation rate led to a change in absorption. What caused the observed change in energy generation remains the main mystery as does the nature of AGN variability in general.

We interpret the *K*-band emission as being due to re-radiation by dust while the *J*- and *H*-band emission is due to a combination of radiation from the outer accretion disk and re-radiation by the hot dust. We have proposed that the dusty clouds are located in a hollow bi-conical outflow of which we see only the near side (Oknyansky et al., 2015; 2017). We have noted that this model can explain cases of changing Seyfert type. Gaskell & Harrington (2017) have proposed that changes in Balmer line profiles such as observed in NGC 2617 are due to clumps in the dusty outflow partially blocking the broad-line region.

We do not think that a tidal disruption event caused by a star passing too close to a massive black hole can be a common mechanism for CL AGNs, since the rate of tidal disruptions is very low. An alternative possibility is that a star does not get close enough to the black hole to be totally disrupted and a less dramatic event might happen. If the stellar orbit is bound and highly eccentric, just like some stars in the centre of our own Galaxy, repeated flares should occur. (Campana et al., 2015; Ivanov and Chernyakova, 2006).

## 6. Conclusions

We have assembled optical light curves of NGC 2617 from 2010 from the MASTER Global Robotic Network, and optical, UV, and X-ray light curves from 2013 onwards based on *Swift* archival data. We have obtained recent spectroscopic and photometric (from NIR toX-ray) monitoring of NGC 2617 in 2016-2017.

We find that NGC 2617 remains in a high state (i.e., it appears as a Seyfert 1), but from the end of the March till the end of July 2017 the object was in very low level of brightness and variability. We suspect that it was then in a transition state and could soon be changing its type back to Sy 1.8.

We continue to find that light-travel-time delays increase with wavelength. The new IR data obtained from October 2016 to May 2017 confirm our published (2016) result of a delay of the *K* band of ~ 24 days relative to UV/optical.

*Acknowledgements.* Thanks to P.B. Ivanov for useful discussions, thanks to the *Swift*, WIRO and MASTER teams for organizing the observations, to A.M. Cherepashchuk and N.S. Dzhalilov for supporting our observations. This work was supported in part by the Russian Foundation for Basic Research through grant 17-52-80139 BRICS-a and by the BRICS Multilateral Joint Sci-ence and Technology Research Collaboration grant #110480. MASTER and ASTRONIRCAM works were supported in part by the M.V.Lomonosov Moscow State University Development Program (equipment). MASTER observations were supported throw RSF grant 16-12-00085 and by the National Research Foundation of South Africa (MASTER-SAAO).